\documentclass[prd,12pt]{article}

\usepackage{amsmath,amssymb,graphicx,multirow,xspace,slashed}
\usepackage[colorlinks=true,urlcolor=blue,anchorcolor=blue,citecolor=blue,filecolor=blue,linkcolor=blue,menucolor=blue,pagecolor=blue]{hyperref}
\usepackage[compress,numbers]{natbib}
\usepackage{subfigure, placeins}
\usepackage{booktabs}
\usepackage{blindtext, rotating}
\usepackage{afterpage}
\usepackage{enumitem}
\usepackage{ marvosym }
\usepackage{authblk} 
\usepackage{verbatim}
\usepackage{soul} 
\usepackage[normalem]{ulem}
\usepackage{pifont}
\usepackage{color}
\usepackage{listings}
\usepackage{setspace}
\usepackage{changepage}

\usepackage[utf8]{inputenc}

\usepackage{floatrow}
\newfloatcommand{capbtabbox}{table}[][\FBwidth]

\usepackage[font=footnotesize,labelfont=bf]{caption}

\usepackage{lineno}

\allowdisplaybreaks

\long\def\symbolfootnote[#1]#2{\begingroup%
\def\thefootnote{\fnsymbol{footnote}}\footnote[#1]{#2}\endgroup}

\newcommand{\newc}{\newcommand}
\newc{\gsim}{\lower.7ex\hbox{$\;\stackrel{\textstyle>}{\sim}\;$}}
\newc{\lsim}{\lower.7ex\hbox{$\;\stackrel{\textstyle<}{\sim}\;$}}
\newc{\gev}{\,{\rm GeV}}
\newc{\mev}{\,{\rm MeV}}
\newc{\ev}{\,{\rm eV}}
\newc{\kev}{\,{\rm keV}}
\newc{\tev}{\,{\rm TeV}}
\newc{\MHT}{$H_T^{\text{miss}}$}
\newc{\MET}{$\slashed{E}_T$}
\newc{\MTT}{$M_{T2}$}

\newc{\mz}{M_Z}
\newc{\mpl}{M_*}
\newc{\mw}{m_{\rm weak}}
\newc{\nr}[1]{N^c_R{}_{#1}}

\def\beq{\begin{equation}}
\def\eeq{\end{equation}}
\newcommand{\bea}{\begin{eqnarray}\begin{aligned}}
\newcommand{\eea}{\end{aligned}\end{eqnarray}}
\def\bitem{\begin{itemize}}
\def\eitem{\end{itemize}}

\numberwithin{equation}{section}

\newcommand\fverb{\setbox\fverbbox=\hbox\bgroup\verb}

\newbox\fverbbox

\usepackage[usenames,dvipsnames]{xcolor}

\newcommand{\be}{\begin{equation}}
\newcommand{\ee}{\end{equation}}

\newcommand{\et}{E_{10}}
\newcommand{\eh}{E_{100}}

\definecolor{Code}{rgb}{0,0,0}
\definecolor{Decorators}{rgb}{0.5,0.5,0.5}
\definecolor{Numbers}{rgb}{0.5,0,0}
\definecolor{MatchingBrackets}{rgb}{0.25,0.5,0.5}
\definecolor{Keywords}{rgb}{0,0,1}
\definecolor{self}{rgb}{0,0,0}
\definecolor{Strings}{rgb}{0,0.63,0}
\definecolor{Comments}{rgb}{0,0.63,1}
\definecolor{Backquotes}{rgb}{0,0,0}
\definecolor{Classname}{rgb}{0,0,0}
\definecolor{FunctionName}{rgb}{0,0,0}
\definecolor{Operators}{rgb}{0,0,0}
\definecolor{Background}{rgb}{0.98,0.98,0.98}

\lstdefinelanguage{Python}{
numbers=left,
numberstyle=\footnotesize,
numbersep=1em,
xleftmargin=1em,
framextopmargin=2em,
framexbottommargin=2em,
showspaces=false,
showtabs=false,
showstringspaces=false,
frame=l,
tabsize=4,
basicstyle=\ttfamily\small\setstretch{1},
backgroundcolor=\color{Background},
commentstyle=\color{Comments}\slshape,
stringstyle=\color{Strings},
morecomment=[s][\color{Strings}]{"""}{"""},
morecomment=[s][\color{Strings}]{'''}{'''},
morekeywords={import,from,class,def,for,while,if,is,in,elif,else,not,and,or,print,break,continue,return,True,False,None,access,as,,del,except,exec,finally,global,import,lambda,pass,print,raise,try,assert},
keywordstyle={\color{Keywords}\bfseries},
morekeywords={[2]@invariant,pylab,numpy,np,scipy},
keywordstyle={[2]\color{Decorators}\slshape},
emph={self},
emphstyle={\color{self}\slshape},
}

\begin{document}

\baselineskip 0.6cm

\begin{titlepage}

\thispagestyle{empty}
 
\vskip 1cm
    \vspace*{1cm}

\begin{center}

{\Large \bf  Searching for New Physics with Deep Autoencoders}

\vskip 1.0cm
{\large Marco Farina, Yuichiro Nakai, and David Shih }
\vskip 1.0cm
{\it NHETC, Dept.~of Physics and Astronomy\\ Rutgers, The State University of NJ \\ Piscataway, NJ 08854 USA} \\
\vskip 0.5cm
\vskip 1.0cm

\end{center}

\begin{abstract}

We introduce a potentially powerful new method of searching for new physics at the LHC, using autoencoders and unsupervised deep learning. 
The key idea of the autoencoder is that it learns to map ``normal" events back to themselves, but fails to reconstruct ``anomalous" events that it has never encountered before. The reconstruction error can then be used as an anomaly threshold. We demonstrate the effectiveness of this idea using QCD jets as background and boosted top jets and RPV gluino jets as signal. We show that a deep autoencoder can significantly improve signal over background when trained on backgrounds only, or even directly on data which contains a small admixture of signal. Finally we examine the correlation of the autoencoders with jet mass and show how the jet mass distribution can be stable against cuts in 
reconstruction loss. This may be important for estimating QCD backgrounds from data. As a test case we show how one could plausibly discover 400~GeV RPV gluinos using an autoencoder combined with a bump hunt in jet mass. This opens up the exciting possibility of training directly on actual data to discover new physics with no prior expectations or theory prejudice.

\end{abstract}

\end{titlepage}

\setcounter{page}{1}

\vfill\eject

\section{Introduction}
\label{sec:intro}

Deep learning is a hot topic in high energy physics. It has been applied to tagging boosted jets of various kinds  \cite{Cogan:2014oua,Almeida:2015jua,deOliveira:2015xxd,Baldi:2016fql,Barnard:2016qma,Kasieczka:2017nvn,Louppe:2017ipp,Pearkes:2017hku,Datta:2017rhs,Butter:2017cot,Egan:2017ojy,Macaluso:2018tck,Guo:2018hbv,Choi:2018dag,Lim:2018toa}, to quark/gluon discrimination \cite{Komiske:2016rsd,Cheng:2017rdo,Luo:2017ncs}, and full event classification \cite{Bhimji:2017qvb,Nguyen:2018ugw}. These are all examples of supervised learning where the training sets are labeled with truth information. More recently people have been starting to explore forms of weakly-supervised and unsupervised learning (see e.g.\ \cite{Dery:2017fap,Cohen:2017exh,Metodiev:2017vrx,Metodiev:2018ftz,Andreassen:2018apy,Collins:2018epr,DAgnolo:2018cun,Monk:2018zsb,DeSimone:2018efk,Hajer:2018kqm,Pol:2018nhq}). In some weak-supervision approaches, binary classification is attempted on a data sample with only imperfect labels, for instance using class proportions or mixed samples \cite{Dery:2017fap,Cohen:2017exh,Metodiev:2017vrx}. Or there have been recent attempts to train a machine learning algorithm to learn the probability distribution of the background and then compare this to the data to discover new physics \cite{DAgnolo:2018cun,DeSimone:2018efk}. Applications of deep learning in high energy physics do not stop at classification tasks: pile-up removal \cite{Komiske:2017ubm}, generative models \cite{deOliveira:2017pjk} and many others (for a review and more references, we refer to  \cite{Guest:2018yhq}) have all been studied.

Although the LHC has performed hundreds, if not thousands, of searches for new physics since its inception, so far no definitive evidence for physics beyond the Standard Model has turned up. All the searches for new physics in the expected places (supersymmetry, composite Higgs, fourth generations, $Z'$s, etc) have turned up empty. This strongly motivates methods to look for physics without as much top-down theory prejudice. We need more ways to discover the unexpected at the LHC, and here is where unsupervised machine learning comes into play.

\begin{figure}[!t]
  \begin{center}
   \includegraphics[clip, width=16cm]{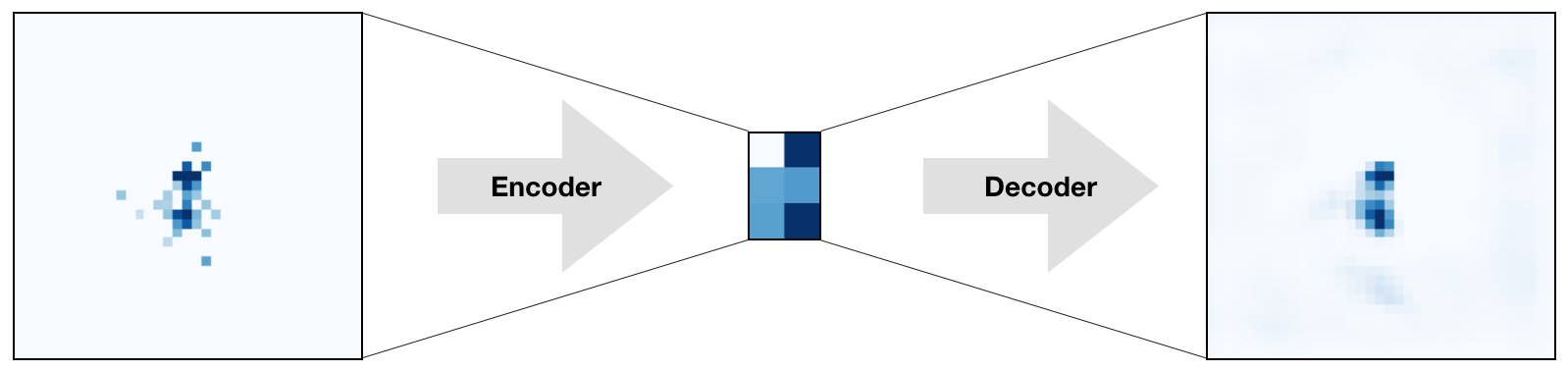}
  \end{center}
  \caption{The schematic diagram of an autoencoder. The input is mapped into a low(er) dimensional representation, in this case 6-dim, and then decoded. }
  \label{fig:simpleautoencoder}
\end{figure}

In this paper, we study one promising avenue to perform open-ended searches for new physics at the LHC: anomaly detection with autoencoders and deep learning. An autoencoder~\cite{BALDI198953}  is a simple idea with various incarnations and many real world applications to anomaly detection, denoising~\cite{Vincent2008ExtractingAC}, generative models~\cite{DBLP:journals/corr/KingmaW13}, feature selection and more. (For an introduction to autoencoders and their applications, see e.g.~\cite{Goodfellow-et-al-2016,DBLP:journals/corr/abs-1801-03149,kerasAE}.)
In its simplest form it is a lossy algorithm that maps an input to a latent compressed representation and then back to itself. This is illustrated in the cartoon in Fig.~\ref{fig:simpleautoencoder}. A measure for how well the autoencoder performs is the difference between input and output according to some distance metric -- the ``reconstruction error". For example, for images, it could be the pixel-wise, summed mean-squared difference between input and output. Typically one trains an autoencoder on a sample of background events with the objective of minimizing reconstruction error on the sample. In this way, it learns what background ``looks like". Any anomaly (the signal, e.g.\ new physics) is then expected to be poorly reconstructed by an autoencoder optimized on a sufficiently different background. Hence we can use a cut on the reconstruction error as an anomaly threshold.

For concreteness, we will focus in this work on distinguishing ``fat" QCD jets from other types of heavier, boosted resonances decaying to jets. Building on previous work on top tagging \cite{Macaluso:2018tck}, we will concentrate on machine learning algorithms that take jet images as inputs. For signal, we will consider all-hadronic top jets, as well as 400 GeV gluinos decaying to 3 jets via RPV.  
Obviously, this is not meant to be an exhaustive study of all possible backgrounds and signals and methods but is just meant to be a proof of concept. The idea of autoencoders for anomaly detection is fully general and not limited to these signals. We will comment on other forms of inputs in section~\ref{sec:discussions}. Moreover there are many other anomaly detection techniques that are not based on autoencoder and/or on reconstruction (loss) which are worth exploring in future work. At the same time autoencoders have been recently used in other high energy physics applications: in parton shower simulation \cite{Monk:2018zsb}, for feature selection of a supervised classification \cite{Hajer:2018kqm}, and for  automated detection of detector aberrations in CMS \cite{Pol:2018nhq}.

We will explore various architectures for the autoencoder, from simple dense neural networks to  convolutional neural networks (CNNs), as well as a shallow linear representation in the form of Principal Component Analysis (PCA). We will see that while they are all effective at improving $S/B$ by factors of $\sim 10$ or more, they have important differences. The reconstruction errors of the dense and PCA autoencoders correlate more highly with jet mass, leading to greater $S/B$ improvement for the 400~GeV gluinos compared to the CNN autoencoder. While this may seem better at first glance, we discuss how one might want to use an autoencoder that is decorrelated with jet mass, in order to obtain data-driven side-band estimates of the QCD background and perform a bump hunt  in jet mass. Indeed, we show how cutting on the reconstruction error of the CNN autoencoder results in stable jet mass distributions, and we show how this can be used to improve $S/B$ by a factor of $\sim 6$ in a jet mass bump hunt for the 400~GeV gluino signal. 

We will study the performance of the autoencoder in two modes: a version where it is trained on background-only events, and a version where it is trained on a mixed sample containing both background and signal, meant to be representative of the actual data. 
An autoencoder trained on a sample of background-only events is an example weakly-supervised machine learning. One could still imagine applying this directly to data, provided one can prepare a control sample that consists only of representative backgrounds. Or one could train on MC backgrounds and hope that the MC is 
an accurate representation of the background events in the data. As a first test of this assumption, we will train on \textsc{Pythia} and evaluate on both \textsc{Pythia} and \textsc{Herwig} and we will see that the results are similar.

By contrast, the autoencoder trained on mixed samples of background and signal is an example of fully-unsupervised machine learning, and as such is a much more exciting potential application. We will show that, surprisingly, the autoencoder performance is remarkably stable against signal contamination: the performance is barely degraded even if signal is 10\% of the training sample! Evidently, there is not much difference between the weakly-supervised and fully-unsupervised modes. Somehow, the autoencoder learns to preferentially reconstruct the background, and still poorly reconstructs the signal, even though it sees the signal as part of the training process. This raises the exciting possibility that the autoencoder could be trained directly on the data, and then could potentially discover any anomalous signal of new physics in the background (perhaps when combined with other variables, for instance a mass cut or bump hunt),  provided it looks different enough from SM objects. This would be an ideal method to discover the unexpected or to perform open ended searches for new physics at the LHC.

Aside from open ended anomaly detection, the autoencoder could be viewed as a general-purpose background-cleaner. That is, we could train it on the background (or directly on the data) and then cut on reconstruction loss in order to remove ``boring" QCD events, leaving behind a sample that is presumably more signal-rich. We could then study these events in more detail, using other techniques and variables to isolate the signal.

The outline of the paper is as follows. In Section~\ref{sec:methods} we define autoencoders quantitatively and present the architectures employed in the rest of the paper. We also describe the details of event generation used to obtain the data sets.  Section~\ref{sec:resultsweak}  is devoted to the main results of the weakly-supervised mode (with pure background training set). We compare the performance of the different architectures, discuss the methods by which we choose the size of the latent space, and perform a MC comparison in the form of \textsc{Pythia} vs.\ \textsc{Herwig}.  In Section~\ref{sec:unsupervised}
 we turn our attention to the fully unsupervised mode. We study the consequences of having a small fraction of signal in the training set, and then we discuss correlation between jet mass and reconstruction loss of the trained autoencoders. We show how by using the CNN autoencoder, a bump hunt in jet mass could potentially reveal the presence of 400~GeV RPV gluinos in the actual data. Finally, we  conclude in Section~\ref{sec:discussions} with a summary and list of future directions.

\section{Methods}
\label{sec:methods}

Let us start with a more detailed introduction to autoencoders. Given an input $x\in \mathbb{R}^n$ we want to learn a mapping into $ \hat{x}\in \mathbb{R}^n$ while passing through a latent representation $y\in \mathbb{R}^k$. This mapping is implemented by two functions: the encoder $f:\mathbb{R}^n\to \mathbb{R}^k$ and the decoder $g:\mathbb{R}^k\to \mathbb{R}^n$. The functional forms of $f$ and $g$ are determined by the autoencoder architecture; they are parametrized by sets of learnable weights, $\theta_f$ and $\theta_g$, respectively.  
 The aim of the autoencoder (and the aim of the machine-learning training process) is to ensure that $x$ and $\hat x=g(f(x;\theta_f);\theta_g)$  are as close as possible under a given metric. Useful results are obtained  when the dimension of the latent space
is smaller than the input one, $ k \ll n$, so that the trivial mapping cannot be learned. Thus the autoencoder learns a compressed representation of the input, optimized on its features.

To evaluate the distance between $x$ and $\hat{x}$ we will use the $L^2$ norm, also known as mean-squared reconstruction error:
\beq
L(x,\hat x) = \frac{1}{n} \sum_{i=1}^n |x_i - \hat{x}_i|^2 \ .
\eeq
By training the autoencoder to minimize $L$ on a sample of background events, we learn to encode and decode the typical events that arise from the background distribution. Then when the autoencoder is evaluated on signals that do not come from the background distribution, the hope is that it will result in a larger $L$ than usual. Thus, the tails of the $L$ distribution are more likely to be signal than background, and by cutting on $L$ we can cut out background and better detect signals. This one of the possible ways to use an autoencoder as anomaly detector.

\subsection{Sample generation}

The jet image samples used in this work follow the exact same specifications as the ``CMS jets" used in \cite{Macaluso:2018tck}. We describe this briefly here but we refer the reader to \cite{Macaluso:2018tck} for more detailed information.

The jets are generated using \textsc{Pythia 8.219} \cite{Sjostrand:2014zea} for hadronization and  \textsc{Delphes~3.4.1} \cite{deFavereau:2013fsa} for detector simulation. All jets are clustered with \textsc{FastJet~3.0.1} \cite{Cacciari:2011ma}. We use anti-$k_T$ jets with $R=1$ and we require $p_T\in [800, 900] \, \rm GeV$ and $|\eta|<1$. 

The background (used for training the autoencoders) consists of light QCD jets, while for examples of signal we will employ top quark jets and gluino jets with mass $m_{\tilde{g}} = 400 \, \rm GeV$. The tops are assumed to decay hadronically, while the gluinos decay to three light-quark jets via RPV SUSY. All the samples are generated by simulating pair production of the heavy resonance starting from $pp$ collisions at 13~TeV (LHC Run II conditions).

In order to ensure that the decay products of the heavy resonance are predominantly contained within the fat jet, we apply a merge requirement of $\Delta R<0.6$ at the truth level on the partonic daughters of the decayed heavy resonance. We also require a geometric match requirement of $\Delta R<0.6$ between the fat jet and the original heavy resonance. 

In all of our studies, we use sample sizes of 100k for training and testing. We have checked with smaller sample sizes that the performance of the autoencoders seems to saturate at 100k, but we have not performed a  detailed study.

After generating the fat jets, we apply several pre-processing steps described in \cite{Macaluso:2018tck} (center, rotate, flip, normalize) and then we pixelize the jets into 37$\times$37 images whose pixel intensities correspond to total $p_T$. We stick to  grayscale images in this work for simplicity.


\subsection{Autoencoder architectures}

In this work, we compare two deep-learning autoencoder architectures, as well as a simpler autoencoder based on principal component analysis (PCA) that could be considered as a baseline. All of our autoencoders take the jet images as inputs. In this subsection we will describe them briefly and qualitatively. In appendix \ref{sec:kerascode}, we will provide full descriptions in the form of Keras code. 

\begin{itemize}

\item For preliminary exploration will use Principal Component Analysis (PCA). The principal components correspond to the eigenvectors of the correlation matrix, ordered in decreasing eigenvalues. The encoder is just a projection on the first $k$ components and the decoder the projection back to the original space. It can be shown that this minimizes the mean-squared error in the space of linear projections. Thus in this sense PCA is comparable to a linear model (e.g. one layer with linear activations and $k$ the dimension of the latent space) with the convenient property of being deterministic.

\item The simplest architecture we consider is just a series of dense (fully-connected) layers. One starts by flattening the $N\times N$ image into a single column vector of length $N^2$. This is then fed to the dense layers of successively smaller size until one arrives at the latent layer. Then this process is reversed until one arrives back at a column vector of the initial size. 

\item For a more sophisticated autoencoder, we consider a convolutional neural network (CNN). Here the dimensionality reduction is accomplished via the usual max-pooling layers. After a series of convolutional and max-pooling layers, the output is fed to a series of dense layers, resulting finally in the latent representation. The entire process is reversed (with 2D upsampling layers in place of the max-pooling layers) to arrive back at an image with the same dimensions. (For the arithmetic of the max-pooling and upsampling to work out, we zero-pad the inputs to the CNN autoencoder so that they are 40$\times$40 pixels.)

\end{itemize}

All the architectures have been implemented using Keras 2.1.5 with Tensorflow 1.7.0 backend on Nvidia GPUs  (Pascal 100 and GeForce GTX 1080). 
For training, we used the default Adam algorithm with minibatch size of 1024\footnote{We found that a smaller minibatch size resulted in worse performance -- the autoencoder converged too quickly and then overtrained.} and a mild early stopping criterion: threshold$=0$ and patience$=3$ ($=5$) for the CNN (dense) autoencoder. As this is a proof-of-concept paper, we have not optimized heavily the training algorithm (e.g.~we have not studied the effect of learning rate annealing or momentum).


\section{Training on backgrounds: weakly supervised mode}
\label{sec:resultsweak}

We now present our results for each autoencoder described in the previous section.
In this section we study the weakly-supervised case with pure background events for training, leaving the unsupervised case with samples contaminated by a small fraction of anomalous events to the next section.

\begin{figure}[t]
\begin{center}
\includegraphics[width=0.7\textwidth]{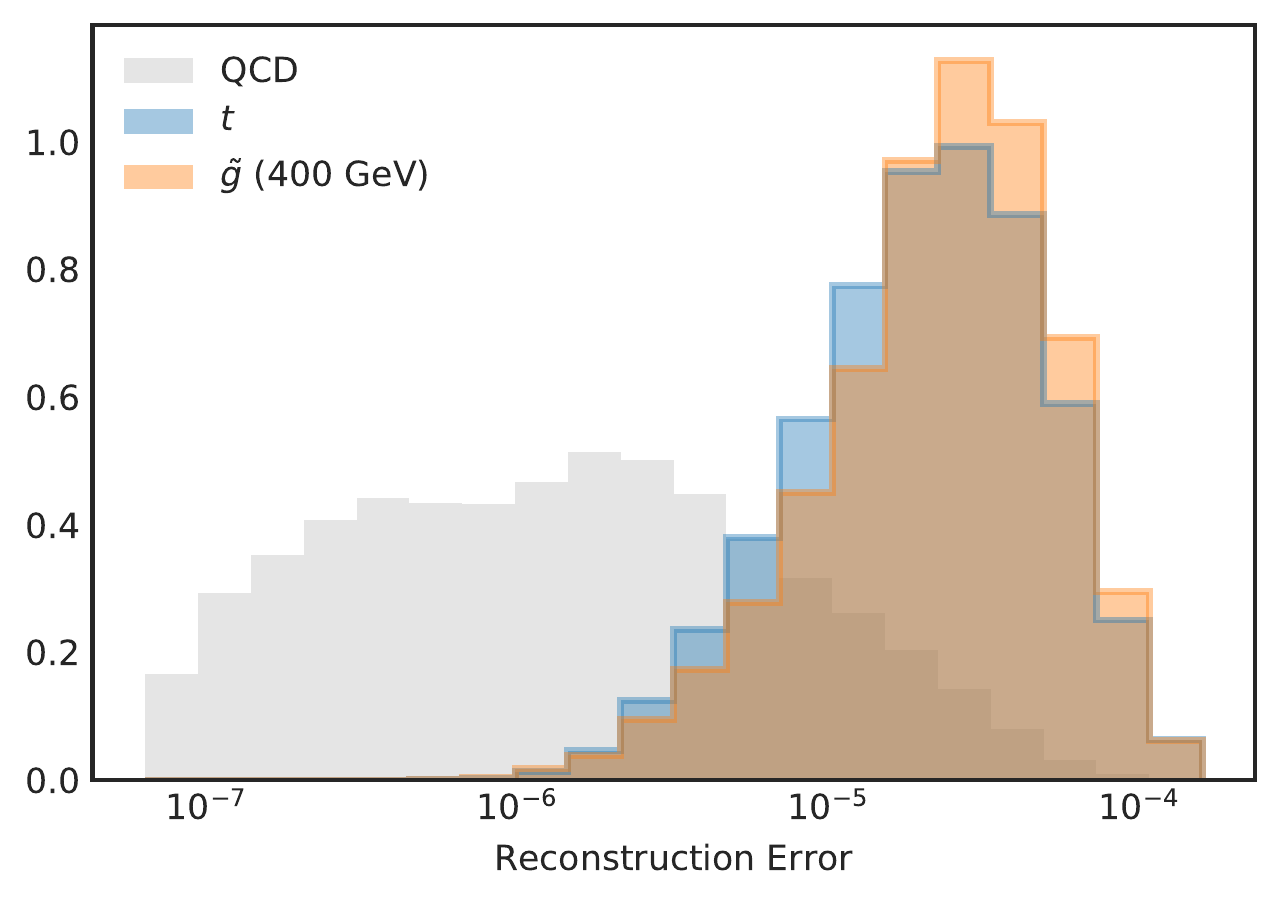}
\end{center}
\vspace{-.3cm}
\caption{ 
Distribution of reconstruction error computed with a CNN autoencoder on test samples of QCD background (gray) and two signals: tops (blue) and $400 \, \rm GeV$ gluinos (orange). }
\label{fig:recolosshistoall}
\end{figure}

\subsection{Autoencoder performance}

\begin{figure}[t]
\begin{center}
\includegraphics[width=0.7\textwidth]{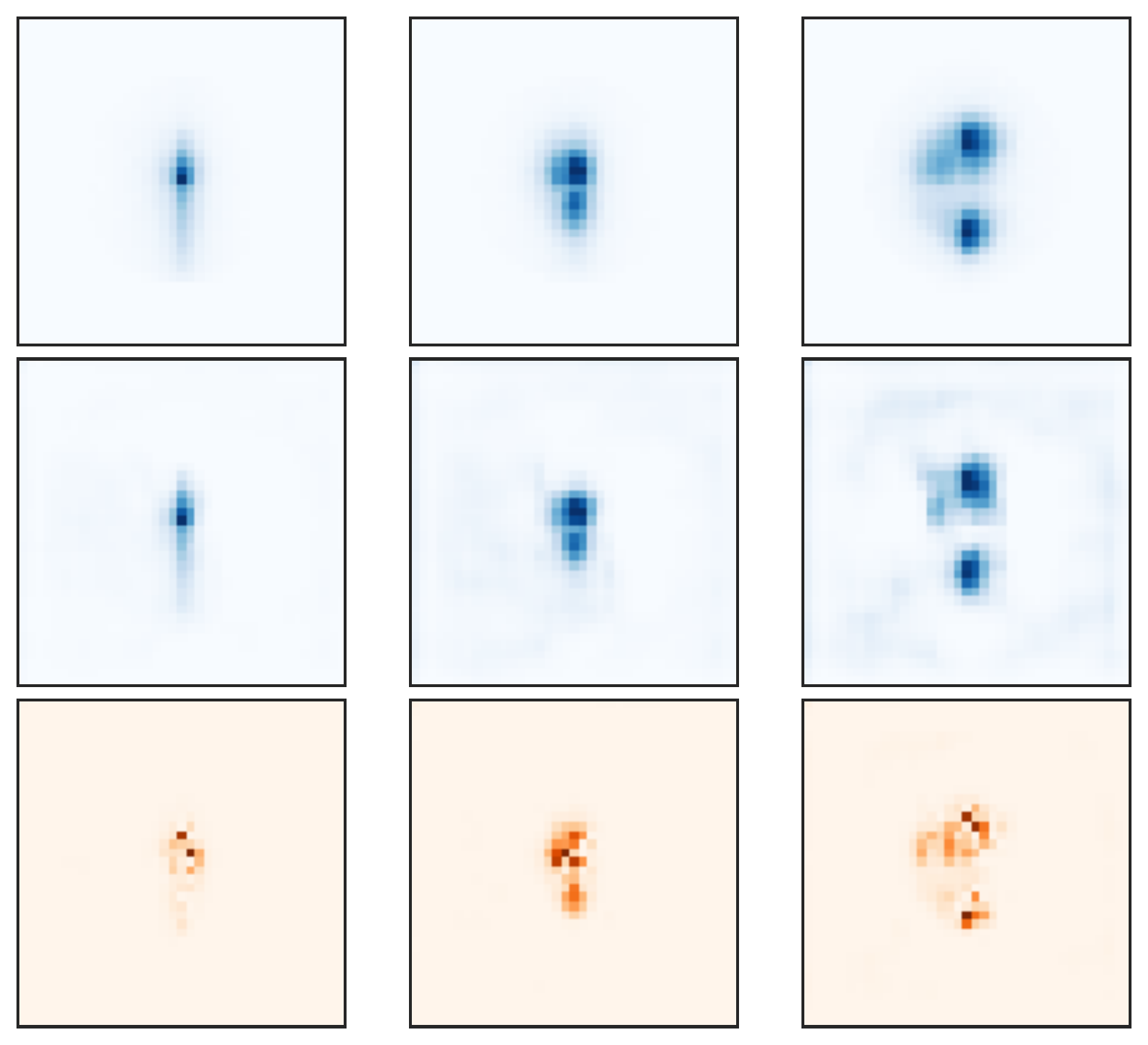}
\end{center}
\vspace{-.3cm}
\caption{ \footnotesize
Each panel represents the average of 100k jet images. Pixel intensity corresponds to the total $p_T$ in each pixel. Upper row: original sample. Middle row: after reconstruction. Lower row: pixel-wise squared error. Left column: QCD jets. Middle column: top jets. Right column: $\tilde{g}$ jets. }
\label{fig:jetavg}
\end{figure}

\begin{figure}[t]
\begin{center}
\includegraphics[width=0.48\textwidth]{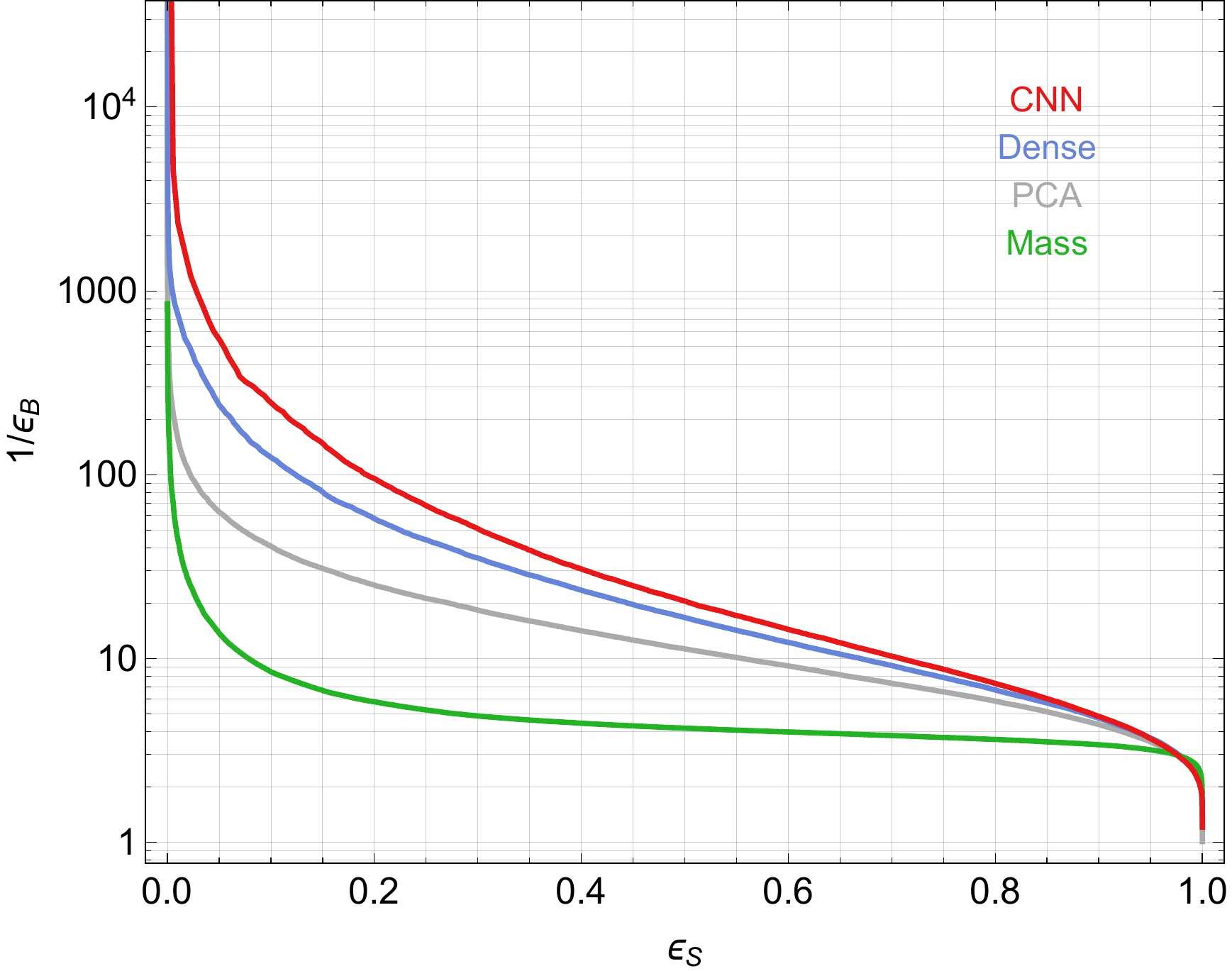}
\includegraphics[width=0.48\textwidth]{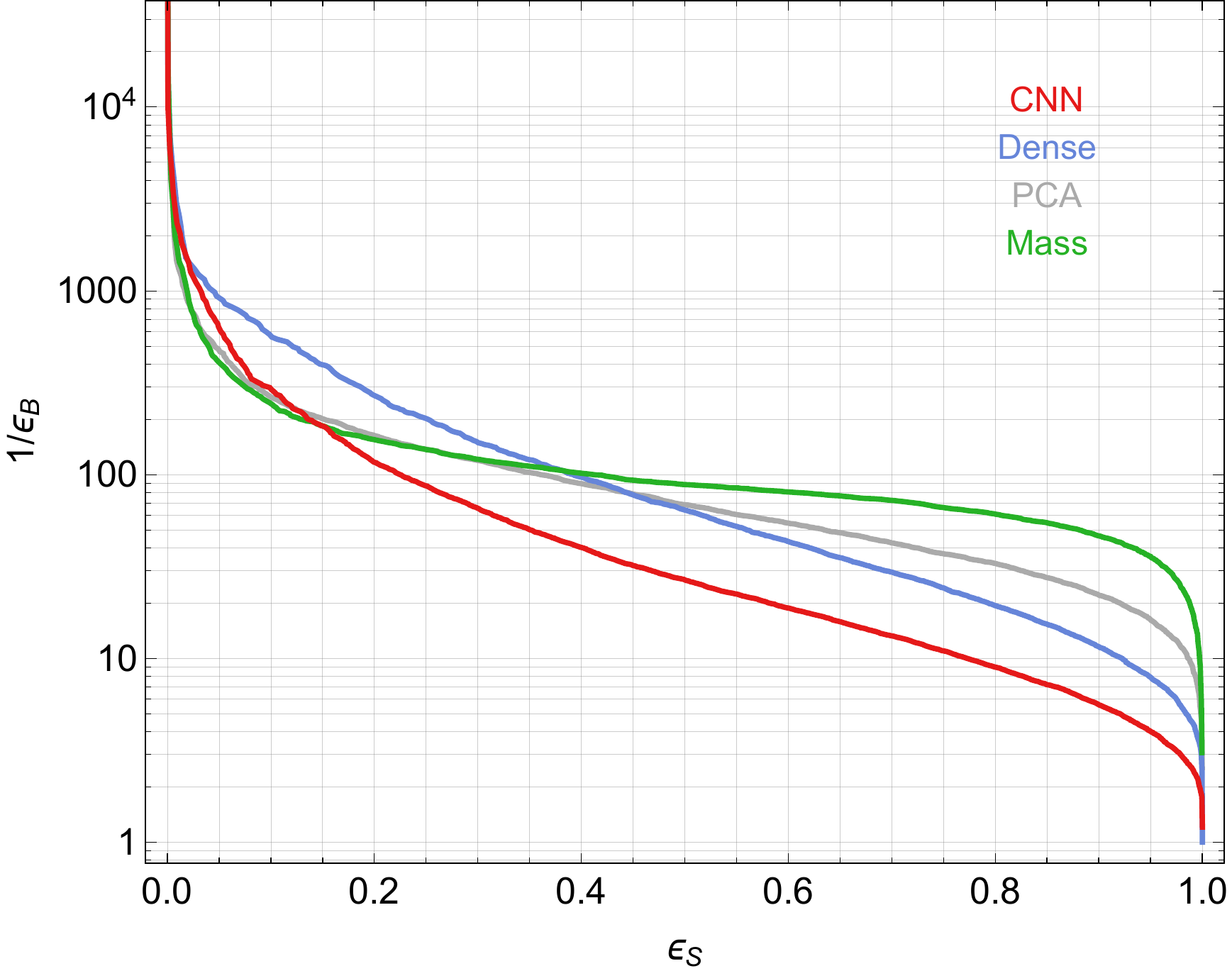}
\end{center}
\vspace{-.3cm}
\caption{ \footnotesize
ROC curves of tagging efficiency $\epsilon_S$ vs background rejection $1/\epsilon_B$ computed on test samples consisting of top jets (left) and gluino jets (right). 
}
\label{fig:roc}
\end{figure}

Shown in Fig.~\ref{fig:recolosshistoall} are histograms of the reconstruction errors for the background sample of QCD jets and the two different signals we consider in this paper (tops and gluinos). We see that the autoencoder  works as advertised: it learns to reconstruct the QCD background that it has been trained on (to be precise, we train on 100k QCD jets and then we evaluate the autoencoder on a separate sample of QCD jets), and it fails to reconstruct the signals that it has never seen before. This is further illustrated in Fig.~\ref{fig:jetavg}, which shows the average QCD, top and gluino jet image before and after autoencoder reconstruction. We see by eye that the QCD images are reconstructed well on average, while the others contain more errors.

By sliding the reconstruction loss threshold $L>L_S$ around, we can turn the histograms in Fig.~\ref{fig:recolosshistoall} into ROC curves. The ROC curves for the different autoencoder architectures are shown in Fig.~\ref{fig:roc} for the top and gluino signals. For comparison we have also included the ROC curve obtained by cutting on jet mass as an anomaly threshold.
While the three architectures have comparable performances it is clear there are some important differences. For tops, the CNN outperforms the others, while for gluinos the situation is largely reversed. Surprisingly, for gluinos, the CNN is even outperformed by the humble PCA autoencoder at all but the lowest signal efficiencies! We will explore this in more detail in section \ref{subsec:jetmass}, but a clue as to what's going on is shown in the comparison of the PCA ROC curve with the jet mass ROC curve. For gluinos, they track each other extremely closely, suggesting that the PCA reconstruction error is highly correlated with jet mass. We will confirm this in section  \ref{subsec:jetmass}. Evidently, the PCA autoencoder (and to a lesser extent the dense autoencoder) has learned to reconstruct the more numerous low mass QCD jets at the expense of the rarer high mass QCD jets. Meanwhile the CNN has learned information that is not as correlated with the mass, e.g.\ details about the jet substructure. 

In Table~\ref{tab:E10E100}, we show the signal efficiency at 90\% and 99\% background rejection (which we refer to as $E_{10}$ and $E_{100}$ respectively).
The values reported in each case are the average over 5 independent training runs to ameliorate the intrinsic variance (apart from PCA which is deterministic). We see that rejecting $99\%$ of background will keep more than 10\% of the signals for both of the deep-learning-based autoencoders.

\begin{table}[!b]
\begin{center}
{
\begin{tabular}{c|c|c} 
 & $t$   &   $\tilde{g}$   \\
 
 \hline
PCA   &  0.51 / 0.04 & 0.98 / 0.36     \\
\hline
Dense & 0.66 / 0.13   & 0.90 / 0.39    \\
\hline
CNN  & 0.70 / 0.19 & 0.77 / 0.23   \\

\end{tabular}}
\end{center}
\caption{$E_{10}$ and $E_{100}$ values for various signals. Results for dense and CNN are obtained as the average of 5 runs of training on the 100k sample (the variances are at the $\sim 0.01$ level). }
\label{tab:E10E100}
\end{table}

\subsection{Choosing the latent dimension}

Here we will explore the dependence of the autoencoder on the dimension of the latent space. This is one of the most important choices to make in the design of an autoencoder for anomaly detection. If the dimensionality is too low, the autoencoder is not able to capture all the salient features of the training set. On the other hand, as the encoding space gets larger, we get closer to the trivial representation. Hence we would like to find an optimal compromise. 

In choosing the latent dimension of the autoencoder, it is important to keep in mind the unsupervised nature of our endeavor. So optimizing the latent dimension using various signals is not the approach we want to take. 

\begin{figure}[t]
\begin{center}
\includegraphics[width=0.95\textwidth]{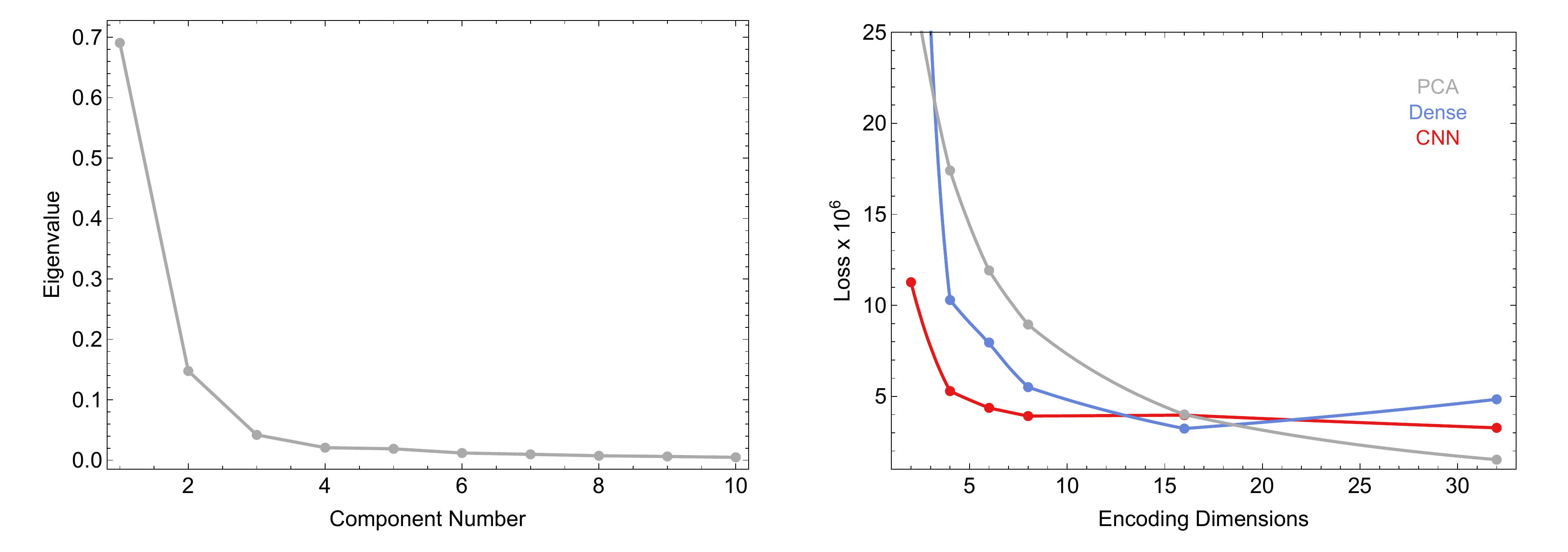}
\end{center}
\vspace{-.3cm}
\caption{ \footnotesize
Left: Scree plot for PCA. Contribution to the variance of each principal component in descending order. Right: average loss as a function of encoding space dimensions.  Each dot corresponds to the average of 5 independent training runs on the 100k training sample (apart from PCA, which is deterministic and has no variance).}
\label{fig:latentdimloss}
\end{figure}

One unsupervised method for finding an optimal working point is to use PCA as the initial step. Shown in Fig.~\ref{fig:latentdimloss} (left) is the amount of variance in the data explained by each eigenvector of PCA, in descending order. (This kind of plot is conventionally referred to as a ``scree plot" by PCA practitioners who also happen to be mountaineers.) An obvious and common prescription is to choose the number of principal components close to the ``elbow" of the scree plot; other choices might be motivated upon more detailed inspection of the cumulative accounted variance (e.g.\ one might choose the number of encoding dimensions corresponding to 95\% or 99\% of the total variance). We could then use the same value for the dimensionality of the encoding space in our deep networks. 

\begin{figure}[t]
\begin{center}
\includegraphics[width=0.95\textwidth]{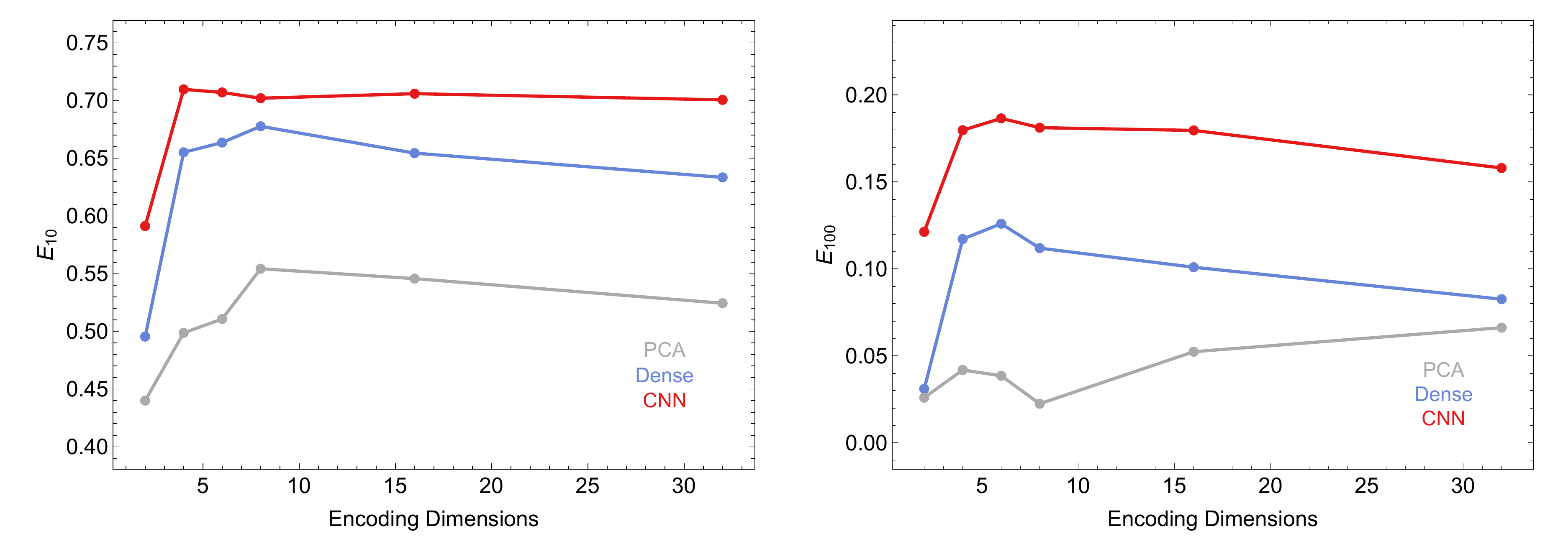}
\end{center}
\vspace{-.3cm}
\caption{ \footnotesize
Dependence of performance of autoencoders in the weakly-supervised learning on number of dimensions of latent space. The values of $E_{10}$ and $E_{100}$ for top jet signals are shown  respectively in the left and right panels.  Each dot corresponds to the average of 5 independent training runs on the 100k training samples (apart from PCA, which is deterministic and has no variance).  }
\label{fig:latentdim}
\end{figure}

We can also search for a similar behaviour in the loss function. This is shown in Fig.~\ref{fig:latentdimloss} (right) for the different autoencoders. We see the loss plateaus around the same place for the various autoencoders, and that corresponds roughly to the elbow of the PCA scree plot. The loss function first sharply decreases as more important and meaningful features are learned by the encoded representation. It reaches a plateau supposedly when only marginal information is added to the encoding space. 
 
 Following the above logic we choose $k=6$ encoding dimensions for all of the autoencoders presented in the paper.
 
Finally, let's examine the wisdom of our choice by looking at the top signal for example. 
 Shown in Fig.~\ref{fig:latentdim} is $\et$ and $\eh$  for the top signal (averaged over 5 training runs) as a function of the latent dimension. This shows the same behavior as we saw above -- the performance of the autoencoders plateau around $k=6$. This is encouraging evidence for our unsupervised method of choosing the latent dimension based on PCA and reconstruction loss.
 
\subsection{Robustness with other Monte Carlo}

Before turning to unsupervised approaches in the next section, let us consider here the main weakness of the weakly-supervised approach: the reliance on accurate background-only samples for training. 

One data-driven approach would be to define a control sample of fat jets, e.g.\ by inverting a lepton selection. This of course assumes the signal is never produced in association with leptons.

Alternatively, one would train on background Monte Carlo, and then apply the autoencoder to data. This would work only insofar as the Monte Carlo accurately represents the background in the data.  Or that any artifacts special to the Monte Carlo are not learned by the autoencoder.
In particular different hadronization schemes could have an impact on the final shape of the jets we study and deteriorate the results of an autoencoder. 

\begin{figure}[b!]
\begin{center}
~~\includegraphics[width=0.8\textwidth]{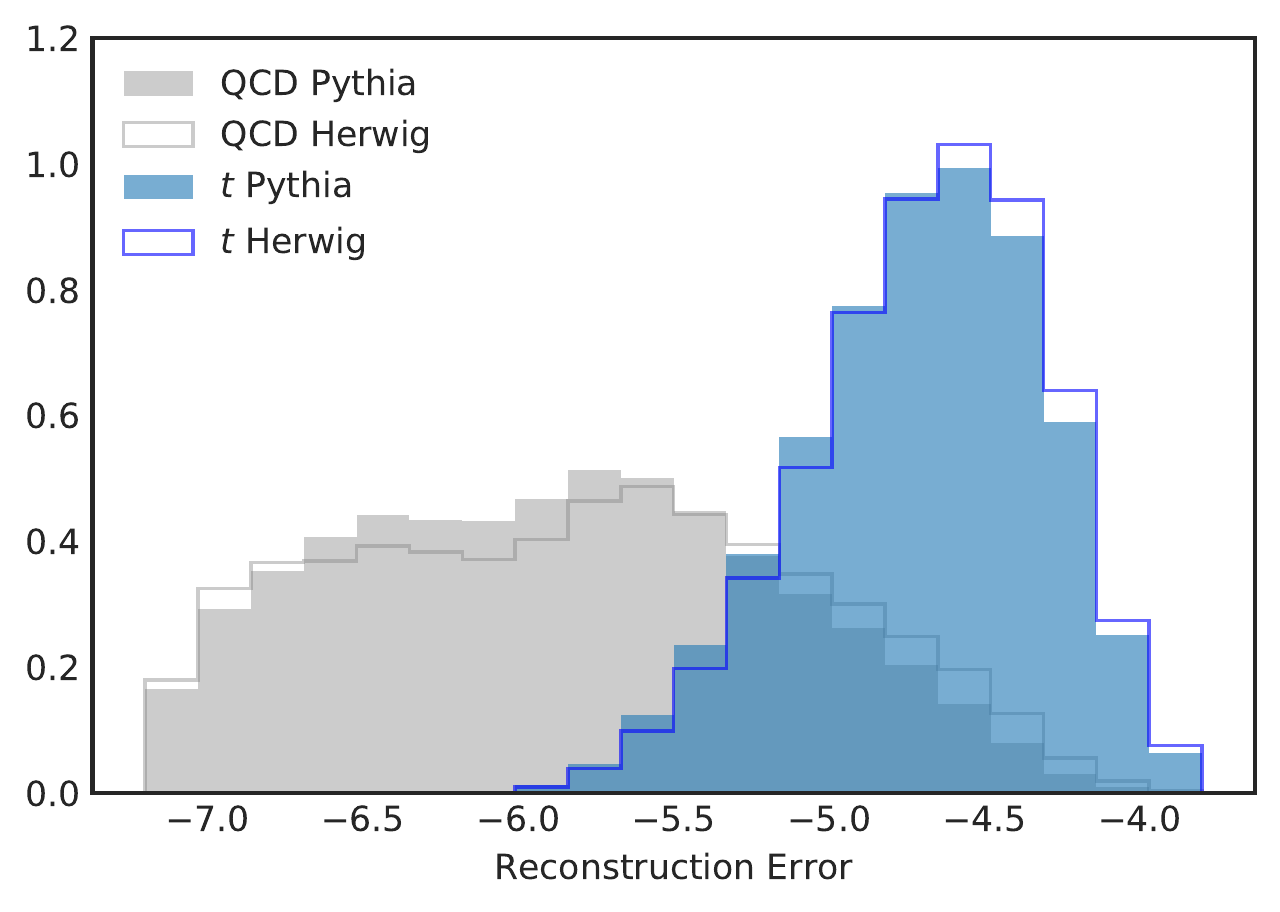}~~
\end{center}
\vspace{-.3cm}
\caption{ \footnotesize
Comparison of reconstruction error distributions between \textsc{Pythia} and \textsc{Herwig} generated test samples, full colored histograms and outlines respectively. Gray is QCD and blue tops. The results are obtained after training a CNN on the \textsc{Pythia} train dataset. }

\label{fig:herwig}
\end{figure}

In this subsection, we will explore the dependence of the autoencoder on the choice of MC generator by evaluating our CNN autoencoder (trained on  \textsc{Pythia}) on fat jets produced with \textsc{Herwig}. Fig.~\ref{fig:herwig} shows the resulting distributions of the reconstruction error. The differences are small, and crucially the separation between background and anomaly is preserved. This can be seen as another proof that the autoencoder has mostly learned fundamental jet features which should depend only weakly on the hadronization scheme details. 

We can quantify the degradation in performance by fixing a common threshold. For convenience we choose it such that on \textsc{Pythia} we have the usual $90\%$ and $99\%$ background rejection. We select one training instance of the CNN autoencoder at random, which corresponds to $\et=0.71$ and $\eh=0.19$. Applying the same threshold and the same algorithm to the \textsc{Herwig} set we obtain precisions of $\epsilon_s=0.74$ and $\epsilon_s=0.21$ respectively, with corresponding background rejection of $87\%$ and $98\%$.

\section{Training directly on data: unsupervised mode}
\label{sec:unsupervised}

\subsection{Contamination study}

In the previous section, we have explored how autoencoders can be trained on samples of background-only jets, and then be used to discover signals such as top quarks and RPV gluinos. This is a prime example of ``one-class classification" and weakly-supervised learning. It could potentially have direct applications to LHC searches for new physics, provided the background sample can be validated somehow. 

In this section, we will turn to a potentially much more exciting application of autoencoders in the form of unsupervised learning. Rather than train on a sample of background-only jets, we will train on a sample of backgrounds ``contaminated" by a small fraction of signal events. We will see how, somewhat surprisingly, the autoencoder still succeeds in detecting anomalies in the test set even though they are present in the training set. Evidently, as long as the autoencoder doesn't see ``too many" anomalies in the course of its training, its performance will be largely preserved.

\begin{figure}[t]
\begin{center}
\includegraphics[width=0.95\textwidth]{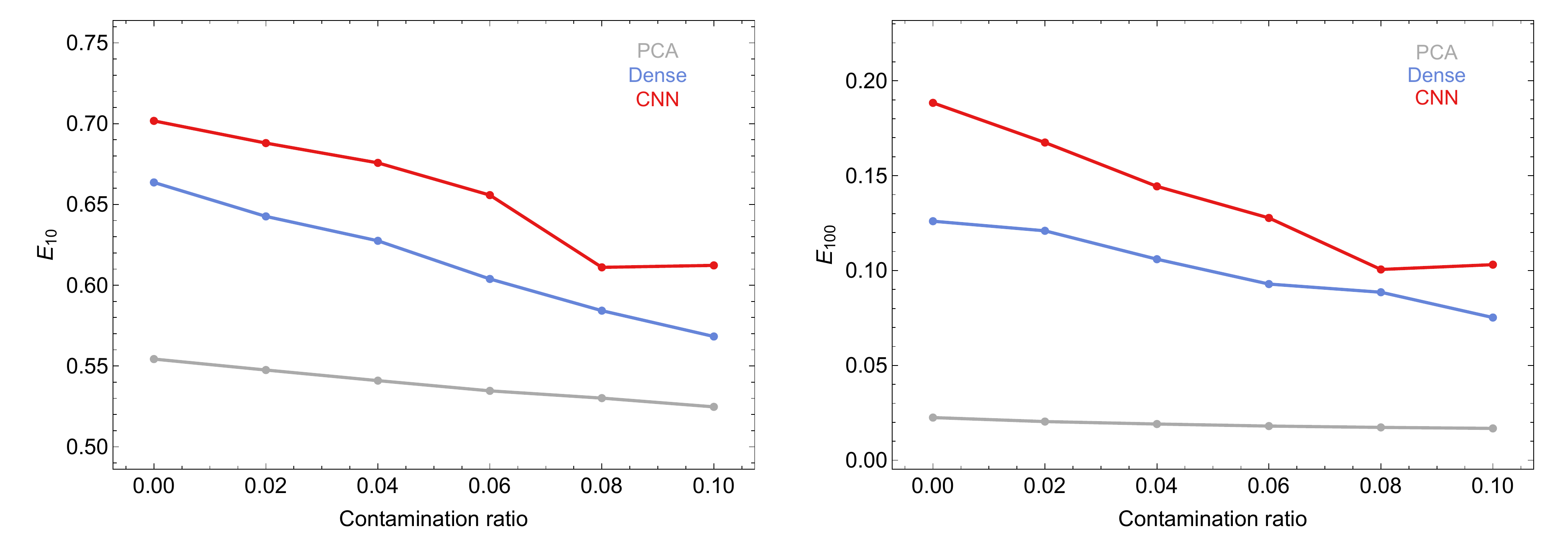}
\end{center}
\vspace{-.3cm}
\caption{ \footnotesize
The performance of autoencoders in the unsupervised learning case where
the training set is contaminated with anomalous events.
We take top jet samples for anomalous events.
The horizontal axis denotes the ratio of top jet samples in the whole training set with 100k samples.
In the left and right panels, the values of $E_{10}$ and $E_{100}$ for top jet signals are shown respectively.
The blue, purple and red curves denote the cases of the simple, 1d and 2d convolutional autoencoders (each dot representing the average of 5 runs), gray for PCA.}
\label{fig:unsupervised}
\end{figure}

Figure~\ref{fig:unsupervised} shows how the amount of contamination with anomalous events in the training set affects the performance of autoencoders.
Here, we use top jet samples for anomalous events.
The horizontal axis denotes the fraction of top jets in the entire training set.
In the left and right panels, the values of $E_{10}$ and $E_{100}$ for top jet signals are shown respectively. For dense and CNN autoencoders, each point represents the average of 5 runs.
In every architecture, as the contamination ratio increases up to 0.1,
the values of $E_{10}$ and $E_{100}$ tend to gradually decrease but the reduction is not dramatic.
This indicates that the contamination does not give a significant impact on the performance of our autoencoders. 

Just to emphasize how powerful this method potentially is, we see that with the CNN autoencoder, even with 10\% signal present in the training sample, the autoencoder arrives at $E_{100}\sim 0.1$, so after this cut on reconstruction loss, we would end up with $S/B\sim {\mathcal O}(1)$!

Of course, without some way of estimating the background, this unsupervised method of searching for new physics would still probably have limited utility.  With just a pure counting experiment (counting the number of events above some reconstruction error threshold), we would have no way of knowing whether we have found new physics, unless we knew beforehand what to expect from the SM background. In the next subsection, we will explore the possibility of combining the autoencoder with a variable like jet mass, in order to perform a bump hunt, with data-driven background estimates coming from sidebands.

\subsection{Correlation with jet mass}
\label{subsec:jetmass}

In this subsection, we will explore the correlation of the different autoencoders with jet mass. We are motivated by how the autoencoder would be applied in the real world to look for new physics.
We are looking for subtle signals in an open-ended way buried in the QCD background. Given that there is no reliable way to estimate the QCD background other than data-driven methods, and given that we are not expecting to achieve extremely high $S/B$ significances, a pure counting experiment seems implausible. Instead, we will still need another variable to side-band in order to estimate the QCD background from the data. Since a large class of new physics starts from the decay of a heavy new resonance, jet mass is an obvious candidate to side band in. 

From this point of a view, the ideal autoencoder would be one whose reconstruction error is minimally correlated with jet mass. We could then cut hard on the reconstruction error to ``clean" out the QCD background, and then look for a bump in the jet mass distribution, confident that the autoencoder cut did not sculpt an artificial peak into the jet mass distribution of the QCD background.

\begin{figure}[t]
\begin{center}
~~\includegraphics[width=0.41\textwidth]{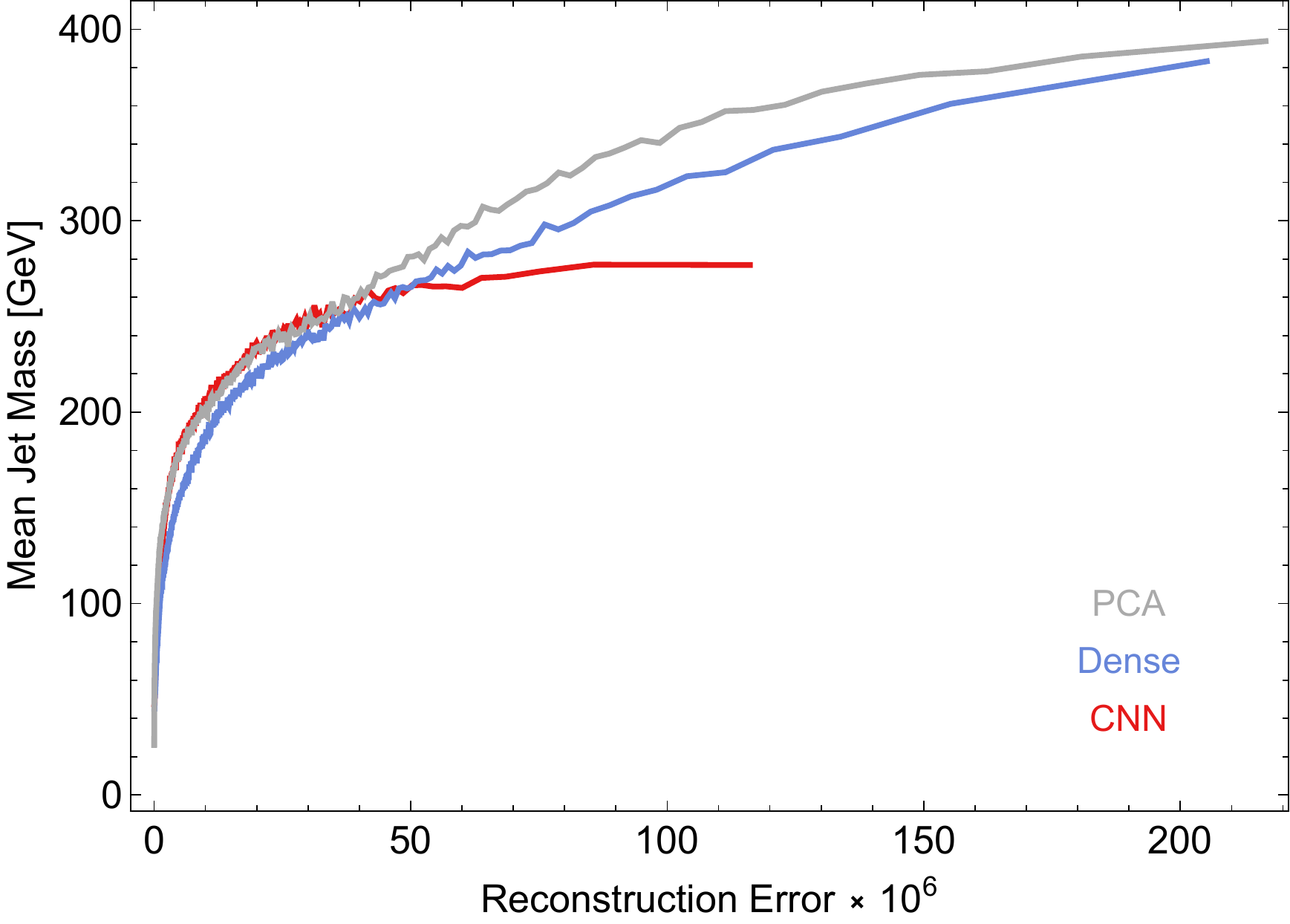}~~~~~\includegraphics[width=0.45\textwidth]{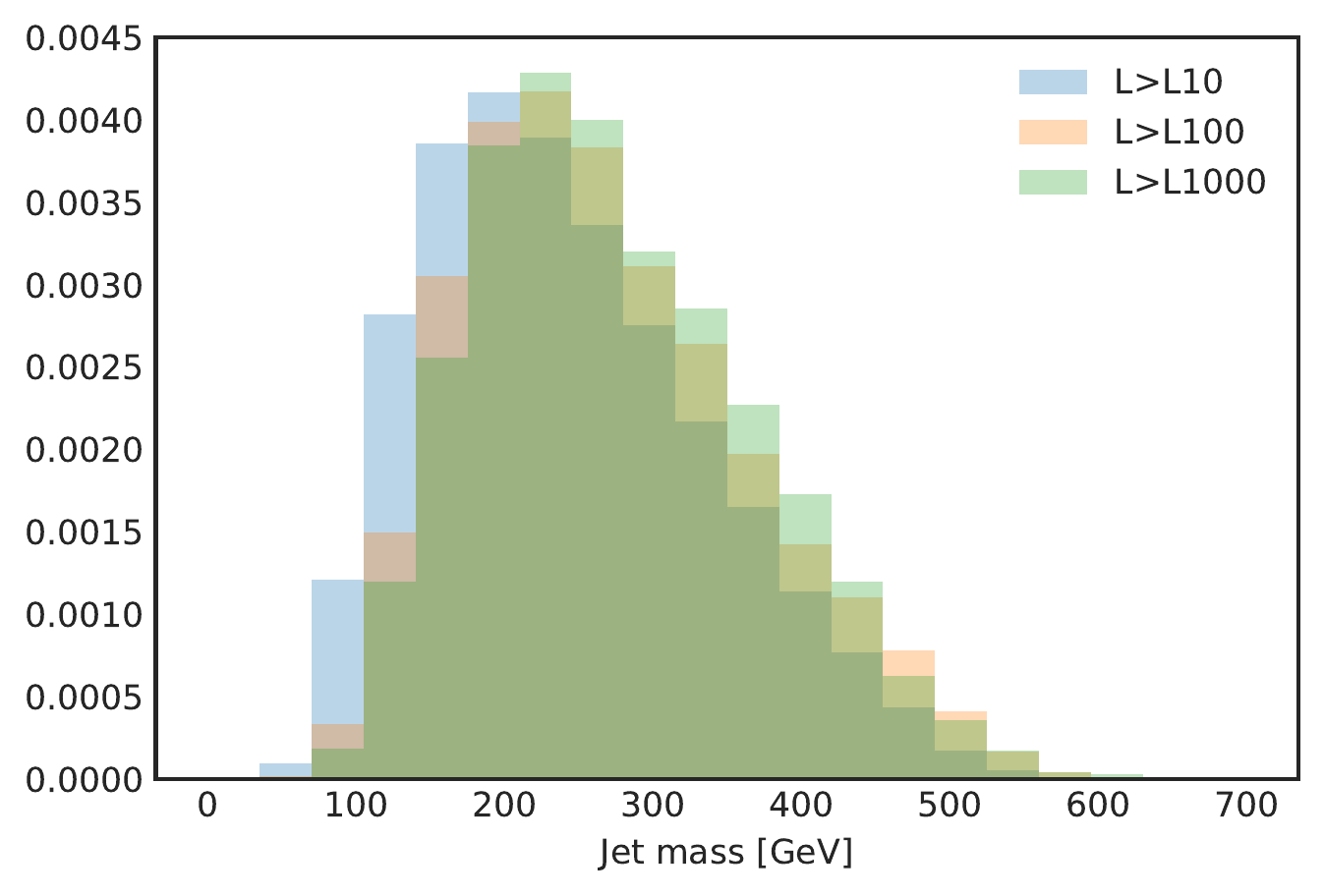}~~
\end{center}
\vspace{-.3cm}
\caption{The left figure shows the average mass in bins of increasing reconstruction error, for the different autoencoder architectures. We see that the PCA and dense autoencoder losses are highly correlated with jet mass all the way up to 400~GeV, while the CNN becomes uncorrelated for masses above $\sim 300$~GeV. The right figure illustrates this with jet mass histograms for the QCD background. We see that they are stable against increasingly hard cuts on the reconstruction error.}
\label{fig:meanmassvsloss}
\end{figure}

Shown in  Fig.~\ref{fig:meanmassvsloss} (left) is the mean jet mass computed in bins of increasing autoencoder loss, for the QCD background. We see that  PCA (gray) and dense (blue) reconstruction errors are correlated with jet mass all the way up to 400~GeV. So cutting on the PCA loss is roughly equivalent to cutting on the jet mass. However, for CNNs the correlation stops for jet masses above $\sim 250$--300~GeV. Equivalently, the jet mass distribution should be stable against cutting on the CNN loss for cuts above $\sim 10^{-6}$. 

This is borne out in Fig.~\ref{fig:meanmassvsloss} (right). Here we see the jet mass distribution after cuts on CNN loss that reduce the QCD background by a factor of 10 (blue), 100 (orange), and 1000 (green). The jet mass distribution is remarkably stable as we cut harder on CNN loss. This makes it the superior autoencoder for doing a bump hunt in jet mass for jet masses above $\sim 300$~GeV.

\begin{figure}[t]
\begin{center}
~~\includegraphics[width=0.48\textwidth]{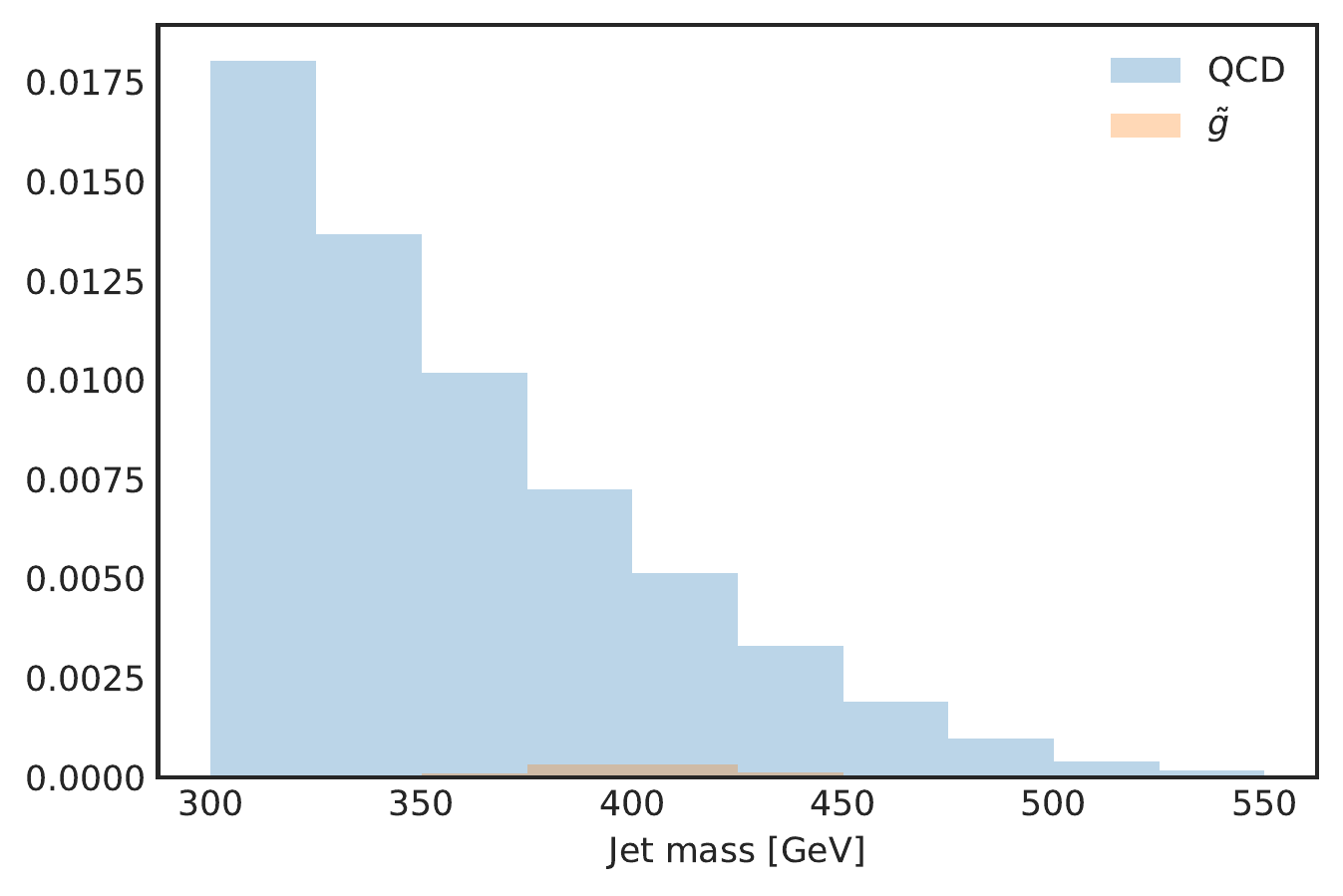}
~~\includegraphics[width=0.48\textwidth]{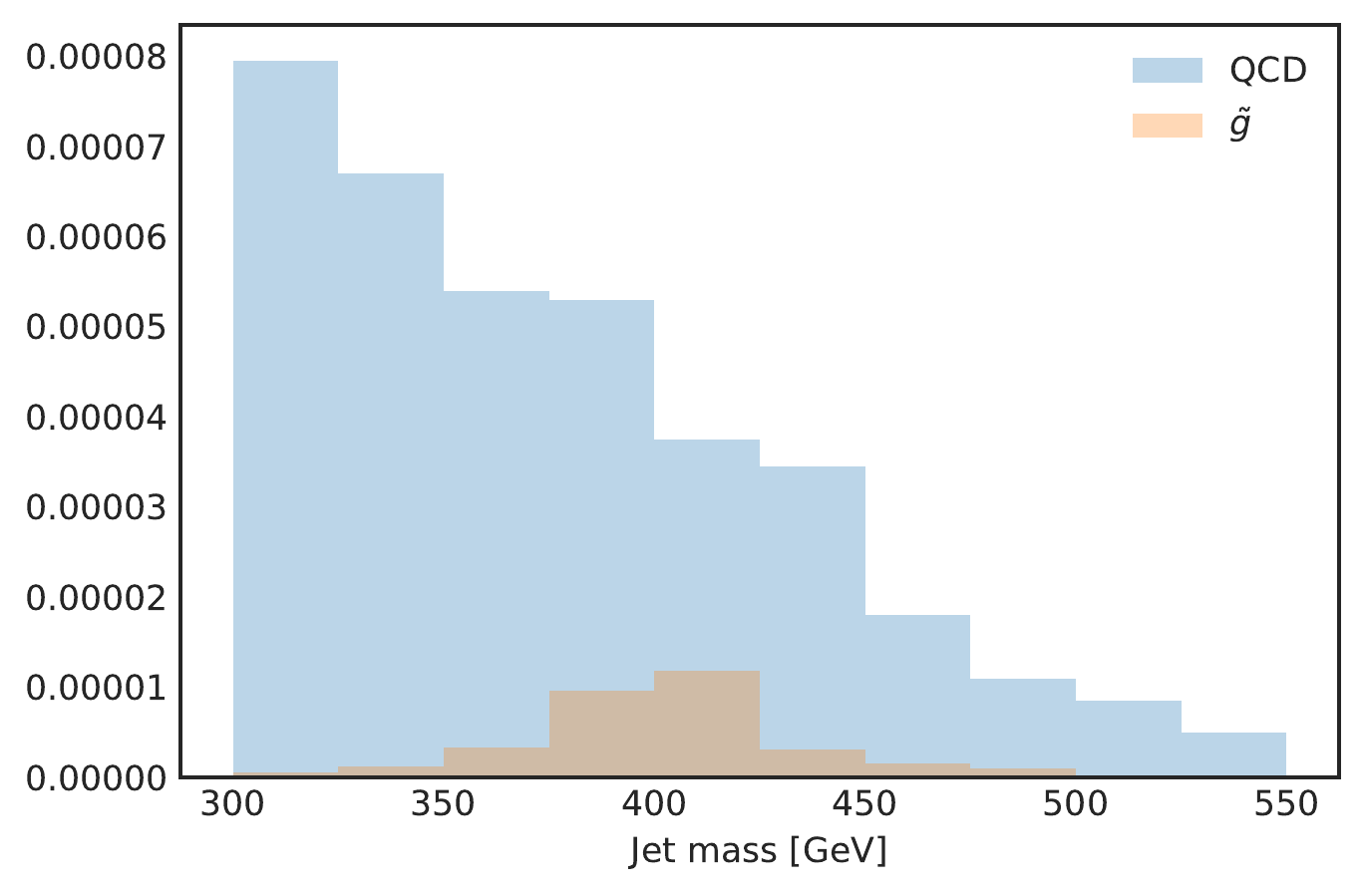}
\end{center}
\vspace{-.3cm}
\caption{Jet mass histograms for QCD background and 400~GeV RPV gluinos, normalized to their LO cross sections, before (left) and after (right) a cut on CNN autoencoder loss that rejects a factor of 1000 of the QCD background. }
\label{fig:gluinomasshistocnn}
\end{figure}

To illustrate the possibilities of searching for new physics in this way, by first ``cleaning" the QCD background using the CNN autoencoder and then doing a bump hunt in jet mass, we include Fig.~\ref{fig:gluinomasshistocnn}. These are the jet mass histograms for QCD background and 400~GeV gluinos, now normalized to the LO gluino and QCD cross sections, before (left) and after (right) a cut on CNN autoencoder loss that removes a factor of 1000 of the QCD background. Importantly, {\it we have trained to autoencoder on a mixed sample containing the expected fraction of gluino jets}, corresponding to a contamination fraction of $10^{-3}$. This would be representative of the actual data, if it really contained these gluinos. We see that the $S/B$ achievable here is $\approx 25\%$. As can be seen clearly from the histograms, this is an impressive improvement on the $S/B$ before the cut (i.e.\ just from the raw jet mass histogram), which is only $\approx 4\%$. One could plausibly discover new physics this way!

\section{Discussion}
\label{sec:discussions}

In this paper, we have shown how autoencoders -- machine-learning algorithms that learn how to compress and decompress a sample of inputs -- are potentially powerful new tools for performing open-ended searches for new physics at the LHC. 
While autoencoders have many real-world applications to anomaly detection, they have up till now not been widely adopted in high energy physics. 

We explored autoencoders in both weakly-supervised and unsupervised forms. In the former mode, we trained autoencoders based on dense and convolutional neural networks on a sample of high $p_T$, $R=1$ QCD jet images and showed how they could learn to accurately reconstruct 
these jet images. Then the hope of using autoencoders for open-ended anomaly detection is that it would fail to reconstruct signals it hadn't been trained on, and then one could use the reconstruction error as an anomaly threshold. In this paper we demonstrated that the deep autoencoders work as advertised, by applying it to signals  consisting of all-hadronic top jets and RPV gluinos. We saw that by thresholding on reconstruction error, the autoencoder 
 could improve $S/B$'s on these signals by sizable amounts.
 
 We also showed how the autoencoder could operate in an unsupervised mode, and discover signals despite having been trained on data that actually contained those signals! In fact, we saw that varying the signal fraction even up to 10\% the autoencoder performance was remarkably stable. This implies that one could simply train the autoencoder directly on the data, and then look for a feature corresponding to new physics. As a proof-of-concept, we showed how this could be done with a jet mass bump hunt. We showed that the CNN autoencoder is reasonably decorrelated with jet mass, meaning that we could use the autoencoder to reduce the QCD background and then search for a bump in the jet mass distribution. We saw that it could achieve $S/B\sim 25\%$ for a 400~GeV RPV gluino signal, an improvement of over a factor of 6 from the bump hunt without autoencoder.

We believe this is a very exciting new direction in the search for new physics at the LHC, very unlike conventional approaches. There are many future directions that we envision. Some of these include:

\begin{itemize}

\item Testing out the autoencoder on other signals and backgrounds. For concreteness, we focused fat jets in a narrow range of $p_T$'s, treating QCD as background and
 heavy resonances with three subjets as signal. But obviously the idea is general and can be applied to any training and test samples in principle.  One could envision applying this to other numbers of subjets, dark showers, non-resonant particles, etc.

\item Going further, it would be fascinating to train an autoencoder to flag entire events as anomalous, instead of just individual fat jets.

\item We focused on just a few autoencoder architectures in this paper, for the proof of principle, but there are many others on the market. For instance, LSTMs and recurrent neural networks. These have proven to be useful for boosted-object tagging 
 \cite{Louppe:2017ipp, Pearkes:2017hku,Egan:2017ojy,Cheng:2017rdo}
so we expect they will also be useful here. There are also even more complex types of anomaly detection in the computer-science literature  based on the idea of  GANs \cite{2014arXiv1406.2661G,DBLP:journals/corr/MakhzaniSJG15,DBLP:journals/corr/SchleglSWSL17} that may also prove useful in this context.

\item It would be interesting to dive deeper into the latent representation that is learned by the autoencoder. Do signals and backgrounds cluster in this latent space? Do the latent dimensions correlate strongly with known variables such as jet mass and N-subjettiness? 

\item We saw here how the CNN autoencoder was reasonably decorrelated with mass. It would be interesting to explore ways to more explicitly decorrelate in mass. The ``variable planing" ideas of \cite{deOliveira:2015xxd,Chang:2017kvc} may be useful in this context. Or one could envision training an 
ensemble of autoencoders on jet samples corresponding to different bins in jet mass. A small enough bin width would probably ensure practical absence of correlation between mass and reconstruction loss. This is well beyond the scope of our study; we reserve this for future work. 
\end{itemize}

 Autoencoders are a form of weakly-supervised or unsupervised machine learning which could be ideally suited to the current situation at the LHC, where many top-down-motivated searches have not turned up any evidence for new physics, and many people are wondering what we should be looking for. With an autoencoder approach, one doesn't need to know what one is looking for. It is a powerful new method to search for any signal of new physics in the data, without prejudice.


\vskip1cm

\noindent{\bf Note added:} While this work was being completed, we learned of the work of \cite{Heimel:2018}, who also studied the applications of autoencoders to anomaly detection and searching for new physics at the LHC. 

\vskip 1cm


\section*{Acknowledgments}
We thank Jack Collins, Kevin Nash, Marc Osherson and Duccio Pappadopulo for helpful discussions.
DS is grateful to the Aspen Center for Physics, which is supported by National Science Foundation grant PHY-1607611, where this work was performed in part.
This work was supported by the DOE grant DE-SC0010008. 

\appendix

\section{Keras code for autoencoder architectures}
\label{sec:kerascode}

\subsection{Dense}

\begin{lstlisting}[language=Python]
input_img = Input(shape=(37*37,))
layer = Dense(32, activation='relu')(input_img)
encoded = Dense(6, activation='relu')(layer)

layer = Dense(32, activation='relu')(encoded)
layer = Dense(37*37, activation='relu')(layer)
decoded=Activation('softmax')(layer)

autoencoder=Model(input_img,decoded)
autoencoder.compile(loss=keras.losses.mean_squared_error,
              optimizer=keras.optimizers.Adam())

\end{lstlisting}

\subsection{CNN}

\begin{lstlisting}[language=Python]
input_img=Input(shape= (40, 40, 1))

layer=input_img
layer=Conv2D(128, kernel_size=(3, 3),
             activation='relu',padding='same')(layer)
layer=MaxPooling2D(pool_size=(2, 2),padding='same')(layer)
layer=Conv2D(128, kernel_size=(3, 3),
             activation='relu',padding='same')(layer)
layer=MaxPooling2D(pool_size=(2, 2),padding='same')(layer)
layer=Conv2D(128, kernel_size=(3, 3),
             activation='relu',padding='same')(layer)
layer=Flatten()(layer)
layer=Dense(32, activation='relu')(layer)
layer=Dense(6)(layer)
encoded=layer

layer=Dense(32, activation='relu')(encoded)
layer=Dense(12800, activation='relu')(layer)
layer=Reshape((10,10,128))(layer)
layer=Conv2D(128, kernel_size=(3, 3),
             activation='relu',padding='same')(layer)
layer=UpSampling2D((2,2))(layer)
layer=Conv2D(128, kernel_size=(3, 3),
             activation='relu',padding='same')(layer)
layer=UpSampling2D((2,2))(layer)
layer=Conv2D(1, kernel_size=(3, 3),padding='same')(layer)    
layer=Reshape((1,1600))(layer)
layer=Activation('softmax')(layer)
decoded=Reshape((40,40,1))(layer)

autoencoder=Model(input_img,decoded)
autoencoder.compile(loss=keras.losses.mean_squared_error,
              optimizer=keras.optimizers.Adam())

\end{lstlisting}

\newpage

\bibliography{ref.bib}
\bibliographystyle{utphys}

\end{document}